\documentstyle[11pt,AATS,epsf]{article}	

\begin{document}	

\title{First Measurement of Uranium/Thorium ratio in a very old star:
Implications for the Age of the Galaxy} 

\author{R.Cayrel,  M. Spite, F. Spite}
\affil{Observatoire de Paris, France}
\author{V. Hill, F. Primas, P. Fran\c{c}ois }
\affil{European Southern Observatory, Germany}
\author{T.C. Beers}
\affil{Michigan State University, USA}
\author{B. Plez}
\affil{GRAAL, Universit\'e Monpellier-2, France}
\author{B. Barbuy}
\affil{Universidade de S\~ao Paulo, Brazil}
\author{J.Andersen, B. Nordstr\"om}
\affil{University of Copenhagen, Denmark}
\author{P. Molaro, P. Bonifacio}
\affil{Osservatorio Astronomico di Trieste, Italy}  


\begin{abstract}
During an ESO-VLT large programme devoted to high reso-lution spectroscopy of
extremely metal-poor stars selected from the HK survey of Beers and colleagues,
the [Fe/H] $= -2.9$ giant CS 31082-001 was found to be as enriched in
neutron-capture r-process elements as CS 22892-052, but with a much-reduced
masking by molecular lines. This allowed the detection and
measurement of the uranium line at 3859 \AA\  for the first time in an
extremely old star.  By making use of the relatively short $^{238}$U decay
(half-life 4.47 Gyr) we obtain a radioactive dating of the formation of the
r-process elements in this star, born in the early days of the Galaxy.
\end{abstract}


\section{Introduction}
Radioactive decay provides a very direct and accurate measurement of time
between the epoch of formation and the present, if one also has a means of
estimating the abundance ratio of the radioactive element relative to a stable
element (or to another radioactive element having a substantially different
half-life).  So far only the ratio of the abundance of $^{232}$Th to a stable
element, usually Eu (also primarily produced by the rapid neutron capture
process), has been used as cosmo-chronometer.  We report here the 
measurement of the abundance of $^{238}$U, which has the great advantage of
having a smaller half-life than $^{232}$Th (4.47 Gyr instead of 14.05 Gyr).
This measurement was made in CS 31082-001, an extremely metal-poor star ([Fe/H]
$= -2.9$) selected from the HK survey of Beers and collaborators (see Beers
1999 for a summary). This star is more metal-deficient than the globular
clusters, and was likely born in the very early Galaxy.  This
measurement, reported in a letter to Nature (Cayrel et al. 2001, issue of
February 8, 2001), was made with the ESO/VLT unit 2 (Kuyen) telescope, and the
UVES spectrograph.  In this presentation I shall concentrate on a discussion of
the abundances of U and Th in CS 31082-001, and  the consequences for
radioactive dating of the material in this star.  In a contribution  by Hill et
al. (this volume) the abundances of the other neutron-capture elements in CS
31082-001 are discussed.  In another contribution Toenjes et al. (this volume)
present computations of the expected production ratios of U/Th, U/Eu, U/Ir,
Th/Eu, and others.  Christlieb et al. (this volume) discuss a strategy for
future detection of additional r-process enhanced metal-poor stars that might
be used  as cosmo-chronometers (see also the discussion by Sneden et al., this
volume).

\section{The Spectroscopic Observation OF CS 31082-001 and Comparison with CS
22892-052}

High-resolution VLT/UVES spectra, in the region of the U II line at 3859.57
\AA, are shown in figures 1 and 2, respectively, for the star CS 22892-052,
already known to be greatly enriched in r-process neutron-capture elements
(Sneden et al. 1996), and the newly discovered star CS 31082-001.

It is quite clear from these spectra,
that both stars are considerably enriched in r-process elements -- the
primary difference between the two stars is the much reduced contamination of
atomic lines by molecular features (mainly CH and CN) in CS 31082-001.  The
visibility of the U II line, excellent in CS 31082-001, is spoiled in CS
22892-052 by the presence of the CN line at 3859.67 \AA\ .  The alreday high  
signal-to-noise ratios of these spectra are  shown here in fig. 1 and
2, enhanced by  a convolution with a Gaussian kernel having FWHM equal to half
the resolution of the spectrograph (only one quarter of the FWHM of the
stellar lines -- producing a negligible loss in spectral resolution).  In the
spectrum of CS 22892-052 the U II line is at the limit of the noise for two
reason: it is blended with the stronger CN line and the line is intrinsically
weaker than in CS 31082-001. Only upper limits to uranium abundance 
have been given so far in CS 22892-052.
   
Another considerable advantage of the lower blending by molecular lines in CS
31082-001 is that we have identified 14 individual lines of thorium in its
spectrum, 10 of which are sufficiently unblended to allow for a precise
determination of its abundance.  In CS 22892-052 there are thus far only 3
lines that might be used for this abundance measurement (Sneden \&
Cowan 2000). Figure 3 illustrates this point (to be compared with fig. 2 of 
Sneden \& Cowan 2000).

\begin{figure}[ht]	
\plotfiddle{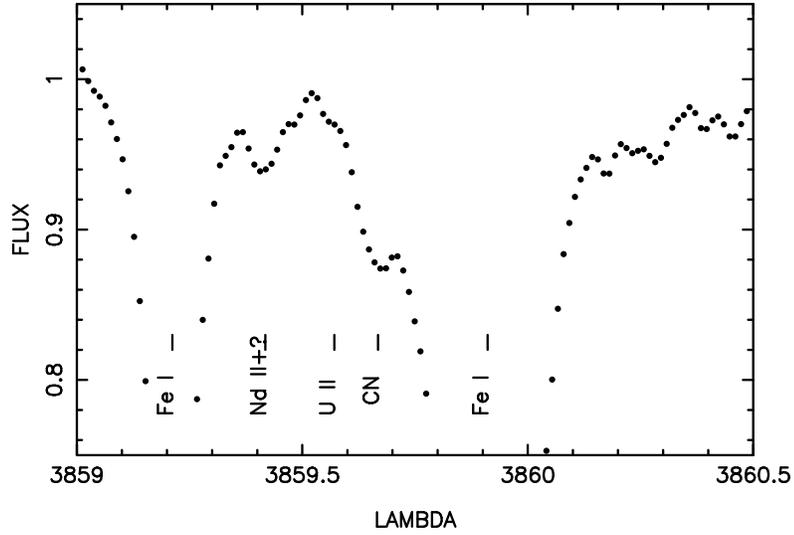}{8cm}{0}{60}{60}{-200}{-20}	
\caption{VLT/UVES spectrum of the star CS 22892-052 in the region of the
uranium line at 3859 \AA}
\end{figure}

\begin{figure}[hb]	
\plotfiddle{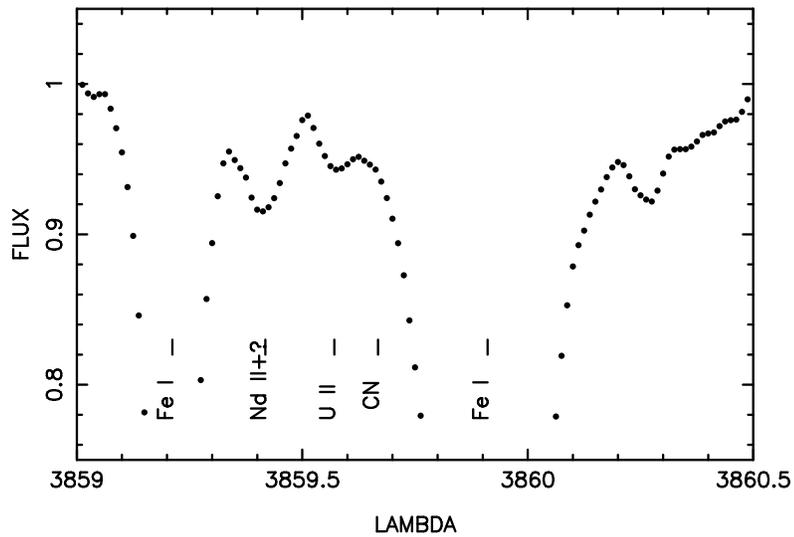}{8cm}{0}{60}{60}{-200}{-20}	
\caption{VLT/UVES spectrum of the star CS 31082-001 in the region of the uranium line
at 3859 \AA}
\end{figure}
\clearpage
\begin{figure}[ht]
\plotfiddle{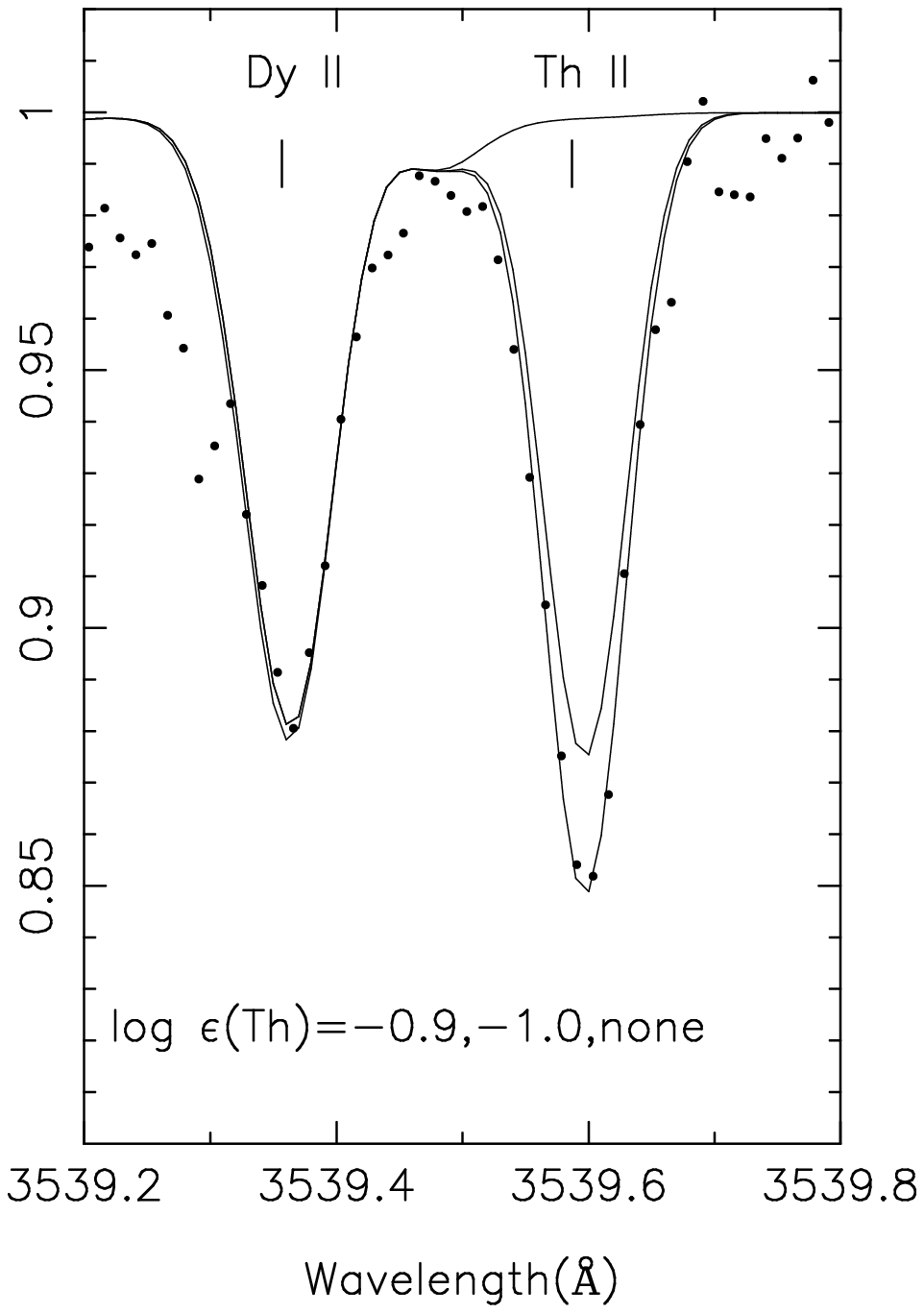}{8cm}{0}{45}{45}{-200}{-24}
\plotfiddle{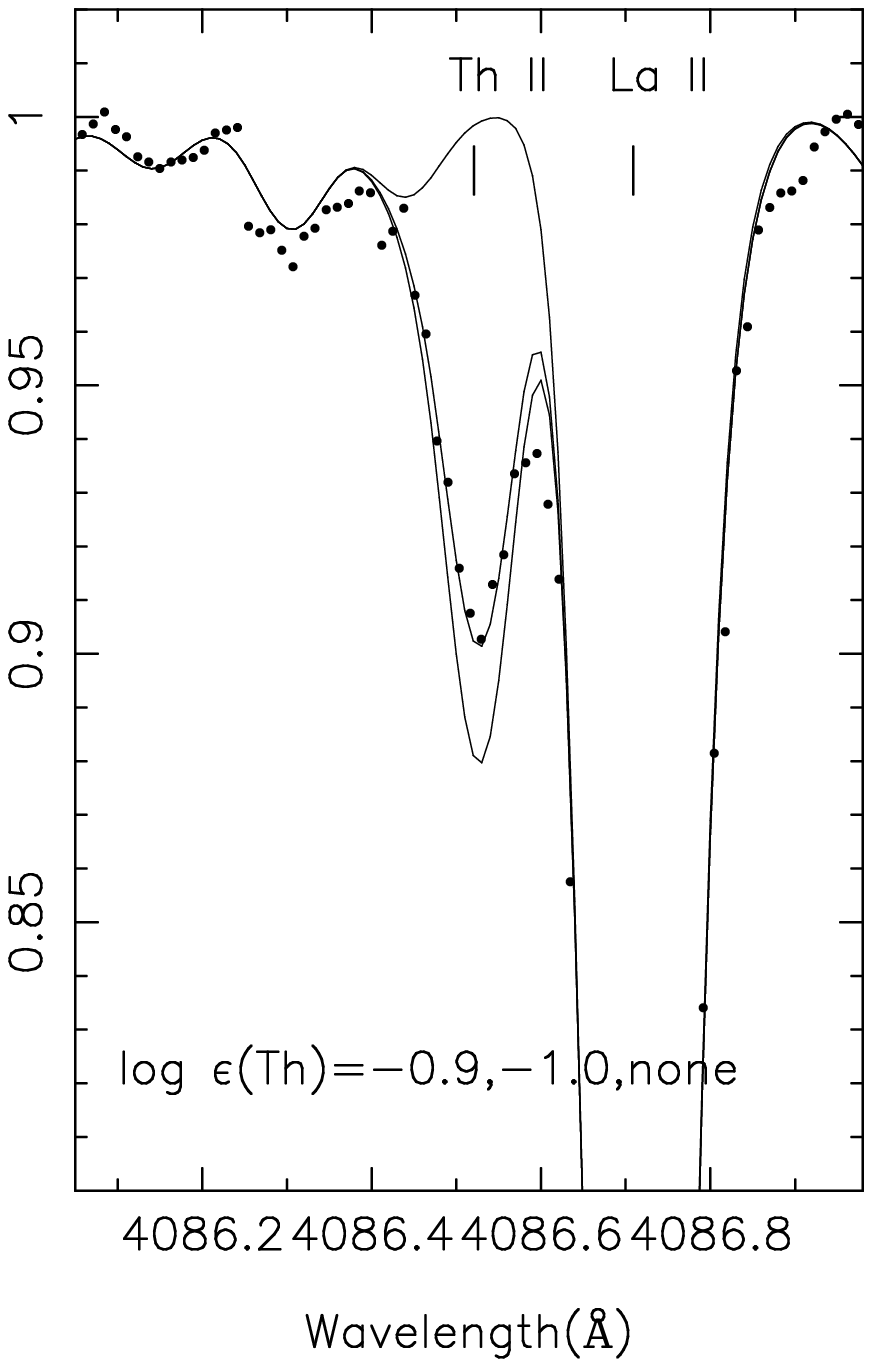}{0cm}{0}{45}{45}{-85}{0}
\plotfiddle{cayrel3c.eps}{0cm}{0}{45}{45}{30}{24}
\caption{Synthesis of three thorium lines in CS\,31082-0012}
\end{figure}

\section{The Abundances of U and Th in CS 31082-001, and the Age Determination}

It has been claimed by Goriely \& Clairbaux (1999) that the ratio U/Th might be
a better cosmo-chronometer than either Th/Eu or Th/Dy, because of the much
smaller mass difference between Th and U than between either one of these two
actinides and the lighter lanthanides (see their fig. 2--5). It is therefore
of particular interest to use the ratio U/Th to determine the age of formation
of these elements, presumably in a type-II SN explosion, the products  of which
were later trapped in the atmosphere of stars such as CS 31082-001, and which
have both decayed with a constant rate, ever since.  Table 1 summarizes the
relationship between the logarithmic concentrations of three chronometer pairs
and time.

\begin{table}		
\begin{center}		
\caption{ Time $\Delta t$ elapsed during epochs at which two abundance ratios
are known, an initial epoch labeled with the subscript  ``0'', and the present
epoch labeled with the subscript ``obs''  \label{table1} }
\vspace*{0.5cm}		
\begin{tabular}{ccc}  	
\tableline		
Time elapsed & & Expression \\                       	
\tableline
$\Delta t$ & = & 46.7[log(Th/r)$_0$ - log(Th/r)$_{obs}$] \\

$\Delta t$ & = & 14.8[log(U/r)$_0$ - log(U/r)$_{obs}$] \\

$\Delta t $ & = & 21.8[log(U/Th)$_0$ - log(U/Th$_{obs}$] \\
\tableline \tableline
\end{tabular}
\end{center}
\end{table}

In Table 1, ``r'' refers to any stable neutron-capture element produced by the
r-process, i.e., with a time between two consecutive neutron captures smaller
than the spontaneous decay of the target.  Because of the large numerical
factor 46.7 in the first expression, resulting from the long half-life of
Th, this chronometer  is less favourable than the other two.  At first
view, the second expression is the best choice.  However, the warning of
Goriely and Clairbaux suggests that it might be quite difficult to reduce the
uncertainty on log(U/r)$_0$, as the first stable r-elements are still far from
U and Th in nuclear mass. Therefore, the last expression is probably the safest
choice.  The increase in the multiplicative coefficient with respect
to the second expression is likely to  be more than compensated for
by the reduction in the uncertainty of the production ratio of log(U/Th)$_0$.

For the present, the best we can do is to select estimates of the
log(U/Th)$_0$ ratio, and derive $ \Delta t$ from expression 3 of Table 1.
In Table 2 we also provide, for comparison, the ages derived from expression 2,
using the heaviest observable stable elements Os and Ir.  The errors do not
include the uncertainties on the theoretical ratios, but do include all other
known sources of  errors, including uncertainties on the oscillator strengths.
Assuming a 0.1 dex uncertainty on log(U/Th)$_0$, and a 0.15 uncertainty on
log(U/Th)$_{obs}$, leads to a global uncertainty of 4 Gyr arising from the use
of U/Th ratio alone.  However, the two independent ratios U/Os and U/Ir
lead to a similar value for the age, with smaller net error bars, because of
the smaller multiplicative coefficient, 14.8 instead of 21.8. We thus consider
3 Gyr to be a reasonable estimate of the global error, taking into account the
the results of the three chronometer pairs.

\begin{table}		
\begin{center}		
\caption{ Ages of r-process elements in CS 31082-001 and their associated
production ratios. The references are (1) Cowan et al. 1999, (2) Goriely \&
Clairbaux 1999 (3) Toenjes et al. 2001 (this volume)
\label{table2} } \vspace*{0.5cm}		
\begin{tabular}{lllll}  	
\tableline		
pair & Log(prod. ratio) & ref  & Log(observ. ratio) & Age \\    
\tableline
 U/Th & $-$0.255 & 1 & $-0.74 \pm 0.15$ & $10.6\pm 3.3$\\

U/Th & $-$0.10 & 2 &  idem & $14.0\pm 3.3$\\

U/Th & $-$0.16  &3 & idem & $ 12.6 \pm 3.3 $ \\

U/Os & $-$1.27 & 1& $-2.19\pm 0.18$ & $13.6 \pm 2.7$\\

U/Ir & $-$1.30 & 1 & $-2.07 \pm 0.17$ & $11.4 \pm 2.5$ \\
\tableline \tableline
\end{tabular}
\end{center}
\end{table}

\section{Relationship Between the Value of (U/Th)$_0$ and the measured ratio in
CS 310822-001 Versus the Solar (U/Th) values}

The present solar value of log(U/Th) is -0.59$\pm 0.05$, according to
Grevesse and Sauval (1998). The value in CS 31082-001 is -0.74$\pm 0.15$ (Table
1). The values at the birth of the Sun are the preceeding ones corrected
by the same factor: the variation of the ratio during 4.6 Gyr, {\it i.e}
0.211 dex , according to table 1, or respectively -0.379 and -0.259, still
differing by 0.15 dex.
 
   If U and Th are always produced in the proportion (U/Th)$_0$,
it is indeed expected that the value of log(U/Th) is smaller in CS 31082-001
than in the Sun, because it has been decaying over the full age of the Galaxy,
whereas in the Sun it has been decaying over any time between 4.6 Gyr and the
age of the Galaxy, depending on its epoch of formation and injection into the
pre-solar nebula.  If we assume that U and Th are formed together continuously
during the time preceeding the birth of the Sun, it is possible to derive
expressions for the build-up of the concentrations $\epsilon _U$ and $\epsilon
_{Th}$ of U and Th in the interstellar medium as:

$$ d\epsilon_U =a_Uf(t)dt $$ $$d\epsilon_{Th} = a_{Th}f(t)dt $$ with: $$
\frac{a_U}{ a_{Th}} = (U/Th)_0 $$ 

\noindent where $f(t)$ is the common history function of the production of
U and Th before the birth of the Sun.  After the birth of the Sun, U/Th has
evolved identically in the Sun and in CS 31082-001, so the difference between
the ratios is only due to the fact that the decay between the origin of the
Galaxy and the time of birth of the Sun $t_{\odot}$ is for CS 31082-001:

	\begin{equation}
	 \frac{\epsilon_U(CS)} {\epsilon _{Th}(CS)} = (U/Th)_0 \exp(-(\alpha_U
-\alpha_{Th})t_{\odot}
\end{equation}

\noindent where $\alpha_U= 0.1551 $ and $\alpha_{Th} =0.04933$.

 For the Sun at the same time $t_{\odot}$:

	$$ \epsilon_{U}(\odot) =
\int_{0}^{t_{\odot}}a_{U}f(t)exp(-\alpha_{U}(t_{\odot}-t))dt $$

	$$ \epsilon_{Th}(\odot) =
\int_{0}^{t_{\odot}}a_{Th}f(t)exp(-\alpha_{Th}(t_{\odot}-t))dt $$

leading to:

\begin{equation}
	\frac{\epsilon _U(\odot)}{  \epsilon _{Th}(\odot)}= 
(U/Th)_0\frac{\int_{0}^{t_{\odot}}f(t)exp( -\alpha_U(t_{\odot}-t)dt}  
{\int_{0}^{t_{\odot}}f(t)exp(-\alpha_{Th}(t_{\odot}-t)dt} 
\end{equation}	

If $f(t)$ or $f(t/t_{\odot})$ is known, the equations (1) and (2)  contain
only two unknown quantities, (U/Th)$_0$ and $t_{\odot}$. It is then possible to
derive (U/Th)$_0$ and $t_{\odot}$, without knowing (U/Th)$_0$ a-priori. 
Actually, this is not our goal, as we prefer to obtain (U/Th)$_0$ from
physicists, and thereby derive as much  information on the astrophysical
unknowns as possible, rather than derive nuclear properties from astrophysical
data.  But at least we can see what a particular value of (U/Th)$_0$ implies
for the history function of production of the r-process elements in the solar
system.  Assuming a uniform production  rate, the result is easily derived.
Taking the ratio of eqn(2) over eqn(1) yields:

\begin{equation}   
\frac{\epsilon_{U}(\odot)/ \epsilon_{Th}(\odot)}{\epsilon_U(CS)/ \epsilon_{Th}(CS)}
= \frac{\int{_0}^{t_{\odot}}exp(\alpha_U
t)dt}{\int_{0}^{t_{\odot}}exp(\alpha_{Th}t)dt}
\end{equation}

The integration is straightforward, and the value of $t_{\odot}$ which gives
the observed ratio of 1.41 (the antilog of 0.15) is 5.9 Gyr.  Adding this to
the 4.6 Gyr elapsed since the birth of the Sun gives an age for CS 31082-001 of
10.5 Gyr.  Plugging the value of $t_{\odot}$ value in eqn(1) gives
(U/Th)$_0$=-0.258, a value almost equal to that listed on the first line of
Table (1). It is equally easy to find the result for $f(t)=\exp(-\lambda t) $.
The value of the parameter $\lambda$ leading to the value of (U/Th)$_0$=-0.16
in the second line of Table 1 is 0.25, giving $t_{\odot}= 8$ Gyr, or as listed
in Table 1, a total age of 12.6 Gyr. This implies a decay of the production
rate of about a factor 7 between the epoch of the early Galaxy and the epoch
of the  birth of the Sun.  This simple calculation emphasizes the
astrophysical impact of the exact value of (U/Th)$_0$.

\section{Conclusion}

Uranium has been detected, and had its abundance measured, in a very
metal-poor star that was born in the early Galaxy, likely before the formation
of the globular clusters.  It has been possible, for the first time, to use
the ratio U/Th for determining the age of formation of these elements in the
early Galaxy.  The accuracy of the result, $12.6\pm 3 $  Gyr, is still rather
severely limited by the 0.15 dex uncertainty on the abundance ratio derived
from observation, and by the 0.1 uncertainty on the theoretical estimation of
the U/Th production ratio.  Progress can be expected in the near future in
three areas: (i) Improvement in the measurement of the oscillator strengths of
U and Th (ii) Improvement in the estimation of the production ratio U/Th$_0$,
based on measurement of the abundances of other heavy elements in CS 31082-001
such as Pb, Bi, Os, Pt, and Ir, as well as  refinements in the nuclear physics 
models, and (iii) Discovery of other metal-deficient r-process enhanced stars
in which U and Th can be measured, to improve the statistics and to provide a
valuable check of the stability of the production ratio in several stars.
   
----------------------------------------------------------------------- %

\acknowledgments
We are indebted to T. von Hippel for scheduling this talk, with great
precision, so  that it took place just at the end of the embargo period on our
letter to Nature.  We are also very indebted to ESO for the generous
allocation of VLT observing time which allowed this discovery to be made.
We thank C.R. Cowley for having called our attention on the fact that uranium
had been observed before, in Ap stars.

\end{document}